\documentclass[preprint]{aastex}	

\newcommand{\HI}{\ion{H}{1}}            
\newcommand{\HII}{\ion{H}{2}}           
\newcommand\etal{{\em et~al.}}          
\newcommand\eg{{\it e.g.}}              
\newcommand\Halpha{{H$\alpha$}}
\newcommand\mjyb{mJy~beam${^{-1}}$}

\newcommand\kms{km~s${^{-1}}$}
\newcommand\Msun{M${_\odot}$}

\renewcommand\deg{{$^\circ$}}

\newcommand\gtsim{\lower.5ex\hbox{$\buildrel > \over \sim$}}
\newcommand\ltsim{\lower.5ex\hbox{$\buildrel < \over \sim$}}

\slugcomment{To appear in the November 2004 issue of the 
{\it Astronomical Journal}}

\usepackage{graphicx}

\shortauthors{Cox \& Sparke}
\shorttitle{Radio Continuum of Polar-Ring Galaxies}

\begin{document}

\title{Radio-Continuum Emission in Polar-Ring Galaxies}

\author{A.~L.~Cox\altaffilmark{1}}
\email{coxa@beloit.edu}
\author{L.~S.~Sparke\altaffilmark{2}}
\altaffiltext{1}{
     Department of Physics \&\ Astronomy,
     Beloit College,
     700 College Street,
     Beloit, WI  53511-5595
     }
\altaffiltext{2}{
     Astronomy Department,
     University of Wisconsin -- Madison,
     475 N. Charter St.,
     Madison, WI 53706
     }


\begin{abstract}
We have used the Very Large Array aperture synthesis telescope to conduct
a radio continuum survey of polar-ring galaxies, at 20\,cm and 6\,cm.
Forty objects were observed at 20\,cm with $\simeq$\,5\arcsec\ resolution.
Twenty (50\%) of the program sources were detected at 20\,cm, down to
our 5-$\sigma$ limit of 0.5~mJy~beam$^{-1}$.  This detection rate is
similar to those in surveys with comparable sensitivity for early-type
galaxies without polar rings.  Sixteen of the objects we detected at
20\,cm were also observed at 6\,cm.  We show radio continuum maps for
the five objects in our sample that have noticeably extended emission.
Our spatial resolution was sufficient to distinguish between emission
originating in the host galaxy from that in the polar ring.  The radio
morphology of the extended sources, as well as the radio to far-infrared
flux ratio and the radio spectral indices of our detected sources,
indicate that star formation, not nuclear activity, is the dominant
source of the radio continuum emission in polar-ring galaxies.  However,
the implied star-formation rates are modest, and only one of our sample
galaxies will consume its supply of cool gas within 500\,Myr.
\end{abstract}



\section{Background \&\ Motivation}
\label{intro}

Polar ring galaxies (PRG's) are early-type galaxies with rings of gas,
stars, and dust that orbit nearly perpendicular to the equatorial plane
of the central galaxy.  It is generally assumed that they result from an
accretion or merger event that occurred some time after the host galaxy
formed \citep[][hereafter PRC]{SWR83, PRC}.  Objects that have undergone
a recent merger often show vigorous star formation (\eg, Schweizer 2000);
hence one might expect PRG's to exhibit signs of enhanced star formation,
either in the polar ring or in the host galaxy, as a result of the
interaction that formed the polar ring.  In many PRG's, the bulk of the
ring material appears to be in regular rotation about the galaxy center
\citep{vGSK87, C96, Aetal97, CSWvM01}, and is presumably stabilized in
some way.  However, some of the accreted ring gas may still be flowing
into the host galaxy; high-resolution optical images give indications
that this may be happening in NGC~4650A \citep{Getal02}.  The inflowing
gas might fuel an active galactic nucleus (AGN) or cause a burst of star
formation in the galaxy center.

Radio continuum emission in galaxies at wavelengths longer than 1\,cm
is generally dominated by synchrotron radiation from ultrarelativistic
electrons.  These cosmic rays may be accelerated either by an active
nucleus or by supernovae of types Ib or II in star-forming regions.
Synchrotron emission from an active nucleus is usually the primary source
of radio continuum emission in early-type galaxies \citep[][hereafter
SJK89]{SJK89}.  In star-forming disk systems, the dominant process
appears to be acceleration by supernovae \citep{Co92}.  Radio emission
is then cospatial with star formation, and free-free radiation from \HII\
regions will contribute at the higher radio frequencies.  It is possible
to distinguish between these two mechanisms for accelerating electrons
and producing radio continuum flux by examining the radio morphology,
the correlation between the radio and far-infrared (FIR) fluxes, and
the radio spectral indices of the sources (see Section~4).

Little information is currently available about the radio/FIR correlation
or radio spectral indices for PRGs.  However, radio continuum emission at
20\,cm has been mapped in a few polar rings to date, often in conjunction
with observations in the 21\,cm line of \HI.  In the polar ring galaxy
ESO~1327-2041 (PRC~C-46; Carilli \&\ \ van~Gorkom 1992) about half
of the emission comes from the optical center of the galaxy, while
the rest is from the brightest knots in the polar ring.  In NGC~4650A
(PRC~A-05; Arnaboldi \etal\ 1997), IC~51 (PRC~B-01; Wilkinson \etal\
1987), and II~Zw~71 (PRC~B-17; Cox \etal\ 2001), the emission is from
the polar material.  The implication is that the radio emission in
these galaxies is from widespread star formation; it is not dominated
by active or star-forming nuclei, perhaps because little of the polar
gas is finding its way to the central regions.

We have used the Very Large Array 
(VLA\footnote{The National Radio Astronomy Observatory is a facility
     of the National Science Foundation operated under cooperative
     agreement by Associated Universities, Inc.})
aperture synthesis telescope to map a sample of PRG's in the 6\,cm
and 20\,cm bands of radio continuum, to measure the total fluxes, and
determine whether the emission comes primarily from the galaxy center
or from the polar material.   We compare our detection rate and radio
spectral indices with those found in radio continuum surveys of early-type
galaxies without polar rings \citep{Sa84, Wetal87, SJK89, Wr91a, Wr91b}.
We also examine the morphology, radio/FIR correlation, and radio spectral
indices for these galaxies, asking whether the emission is likely to
be a consequence of ongoing star formation or an AGN.  Assuming that
star-formation is the dominant source of the radio emission, we derive
star-formation rates and gas-consumption timescales for the survey PRGs.


\section{Sample Selection \&\ Observations}
\label{sample}

We have conducted a survey of 40 PRG's in the 20\,cm and 6\,cm bands,
using the VLA ``B'' and ``C'' configurations, respectively, to obtain
maps with matched resolution ($\simeq$\,5\arcsec) in the two bands.
The objects in our sample were chosen from categories A, B, or C
(confirmed, probable, or possible polar rings) of the ``Polar-Ring
Catalog'' (PRC).  We selected those with $\delta$\,$\geq$\,$-$30\deg\ (far
enough north for a good synthesized beam with the VLA) and angular size
$\leq$\,4\arcmin.  The latter criterion eliminated NGC~660 (PRC~C-13),
a known starburst galaxy \citep{vDetal95}, and NGC~3384 (PRC~C-34).
We also removed the known quasar 3CR~249.1 (PRC~C-36) from our sample.
Fifty-nine galaxies met these criteria.  Forty of these were observed
at 20\,cm, selected only by the hour angles available at our scheduled
observing times.  Sixteen of our 20\,cm targets were also observed at
6\,cm, selecting preferentially those galaxies detected at 20\,cm or with
known redshifts.  The basic data for these galaxies, including positions,
redshifts, and FIR fluxes, are listed in Table~1.

Standard calibration and mapping procedures were followed.  Maps were
made with uniform weighting to maximize spatial resolution.  A source
was counted as detected if it was (1) extended and at least 3-$\sigma$
above the noise at either wavelength, or (2) a point source at least
5-$\sigma$ above the noise at either wavelength, at the position of
the target galaxy.  The probability of detecting an unrelated source
brighter than 1~mJy within 5\arcsec\ of the target galaxy is only 0.05\%\
\citep{Wetal97} --- confusion with background sources should not be a
significant problem at these detection limits.

For the most distant objects in our survey, our detection limit would
allow us to detect any galaxy with an excess of radio emission (larger
than a few times that of a normal spiral galaxy; see \citet{KO88}.
We detected 20 of the 40 observed galaxies at 20\,cm, and 13 of the 16
observed galaxies at 6\,cm.  Our detection rate of 50\% at 20\,cm, down
to a 5-$\sigma$ limiting flux density of $\simeq$\,0.5~mJy~beam$^{-1}$,
is similar to the 42\% detection rate of elliptical galaxies with shells,
down to the same limiting flux density, found by W87.  These detection
rates are not significantly different from the population of ``normal''
early-type galaxies studied by SJK89 (also 42\%) and 
\citet[][hereafter WH91]{WH91}.

The radio continuum fluxes and radio spectral indices for our sample
are shown in Table~2, along with 21\,cm fluxes and \ion{H}{1}\ masses,
when available.  For non-detections, we show 5-$\sigma_{RMS}$ upper
limits, assuming any continuum emission is unresolved.  For unresolved
sources, the total uncertainty was calculated by adding two terms
in quadrature: An assumed 3\%\ uncertainty in the total flux due
to calibration $\sigma_{cal}$ \citep[e.\,g.,][]{WH91}, and the RMS
uncertainty $\sigma_{RMS}$ measured in the CLEANed images.  Typical RMS
uncertainties were 0.1\,\mjyb\ at 20\,cm and 0.08\,\mjyb\ at 6\,cm (with
ranges between 0.06-0.2\,\mjyb\ and 0.03-0.11\,\mjyb, respectively).
For extended sources, the RMS uncertainty in the integrated flux is
increased by a factor of $\sqrt{N}$, where N is the number of beam
areas over which we integrated to measure the source flux.  The total
uncertainty is thus the calibration error $\sigma_{cal}$, added in
quadrature with $\sqrt{N}\sigma_{RMS}$.


\section{Source Morphology}
\label{rcont}

In our sample, 15 (75\%) of the 20 sources we detected are unresolved at
$\simeq$\,5\arcsec\ resolution.  Thirteen of these 15 sources are located
at the optical center of the galaxy, to within our 5\arcsec\ resolution.
One of those thirteen sources, PRC~C-73, has two point sources at both
20\,cm and 6\,cm,  at the positions of the two brightest peaks at optical
wavelengths.  The two unresolved sources for which the radio position does
not correspond to the optical center of the target galaxy are PRC~A-03
(NGC~2685, the ``Helix Galaxy'') and PRC~C-60 (ESO~464-G31).  The radio
source in PRC~A-03 lies 10\arcsec\ from the galaxy center, at a position
angle of 58$^o$, within a region of extended \Halpha\ emission around the
nucleus of the galaxy \citep{EP97}.  Though the ionized gas in this region
does rotate about the minor axis of the host galaxy \citep{S98},
it is not part of what has traditionally been called the polar ring,
which lies at radii beyond $\simeq$\,35\arcsec\ from the galaxy center.
The radio emission in PRC~C-60 coincides with the optical center of
the galaxy MCG--05-50-009, the companion to ESO~464-G31.  These two
galaxies together make up the galaxy pair that was included in the PRC
as a possible polar ring.

The morphology of the extended sources in our sample is quite different
than the radio morphology of typical E/S0 galaxies without polar
rings.  None of the five sources that have extended structure in our
$\simeq$\,5\arcsec\ resolution maps show the ``core-jet'' structure
typical of AGN's.  The radio continuum morphologies are consistent
with star formation either in the candidate polar ring (PRC~B-01 \&\
B-21) or in an extended region of the central galaxy (PRC~C-24, C-28,
\&\ C-51).  Two of the latter sources do have bright peaks at the
galaxy center (PRC~C-28 \&\ C-51), but the rest of the radio emission
in these two galaxies is diffuse and extends throughout the disk of the
central object.  Radio continuum maps for the five extended sources,
as well as for PRC~C-73, are discussed in the following subsections.

\begin{center}
  {\it\bf PRC B-01} (IC~51, Arp~230)
\end{center}

This galaxy is a well-known shell elliptical galaxy, and was included in
the sample of shell elliptical galaxies mapped in the radio continuum
by W87.  It is listed as a candidate polar-ring galaxy in the PRC
because it is known to have a fast-rotating disk or annulus of gas,
dust, and stars perpendicular to the apparent morphological major axis
of the stellar body.  Observations in the 21\,cm line and \Halpha\ by
D. Schiminovich (1996, private communication) show gas components that
rotate about the minor axis of the dust lane/polar ring, but also emission
well above the plane of the dust lane.  The \HI\ appears to be associated
with the outer shells, while the \Halpha\ emission is in a ``fluffy disk''
around the dust lane.  Because of the interesting morphology of the radio
continuum in this galaxy, we observed it for 30 minutes in each of the
VLA B and C configurations.  Thus, we have slightly better sensitivity
on this source than on the other sources in our sample (see Table~2).

The majority of the radio continuum emission in this galaxy, at
both 20\,cm and 6\,cm, is aligned with the dust lane (the candidate
polar ring).  However, with our improved sensitivity, we see filaments
that extend well above the optical plane of the candidate polar ring;
maps made with both uniform and natural weighting, at both wavelengths,
are shown in Figure~1.   The morphology of this source indicates that
current star formation is the most likely source of the radio continuum
emission in this galaxy.

\begin{center}
  {\it\bf PRC B-21} (ESO~603-G21)
\end{center}

The radio continuum emission in our 20\,cm and 6\,cm observations is only
resolved along one direction, coincident with the major axis of the dust
lanes observed in the central portions of the candidate polar ring (PRC;
see Figure~2).  However, the optical extent of the candidate polar ring is
$\simeq$\,50\arcsec (PRC), while the 20\,cm and 6\,cm continuum emission
we detect has an extent of only 10\arcsec-15\arcsec.  \citet{Aetal95}
find that most of the emission at 2\,\micron\ ($K$ band) comes from a
plane close to that of the dust lane.  They suggest that this is not a
PRG, but a disk galaxy with abnormally heavy dust extinction and a very
blue bulge.

\begin{center}
  {\it\bf PRC C-24} (UGC~4261)
\end{center}

The 20\,cm emission in this galaxy does not appear to be aligned with
a particular plane; rather, there is a bright central core, surrounded
by an apparently smooth and nearly round ``halo'' of emission.  If the
extension to the northeast of the majority of the emission is real,
it would be interesting, as it lies approximately along the candidate
polar ring in this galaxy (PRC; see Figure~3).  Although there may be
a bright core of emission in the galaxy center, the radio emission does
not exhibit the typical ``core-jet'' morphology of an AGN.

\begin{center}
  {\it\bf PRC C-28} (NGC~2748)
\end{center}

The emission in this galaxy extends nearly the entire length of the galaxy
disk at 20\,cm.  There is a bright central core, but it contains only
$\simeq$\,5\%\ of the total emission; nearly all of the 20\,cm emission
comes from the diffuse component in the galaxy disk.  None of the emission
appears to be associated with the candidate polar ring (see Figure~4); it
is probably due to star formation in the disk of the host galaxy.

\begin{center}
  {\it\bf PRC C-51} (NGC~6286, Arp~293)
\end{center}

The morphology of the 20\,cm emission in this galaxy is very similar
to that of PRC~C-28 -- there is a bright central core, surrounded
by diffuse emission that extends throughout the disk of the galaxy
(see Figure~5).  However, the unresolved core component in this object
contains $\simeq$\,35\%\ of the total 20\,cm emission in this object.
Again, none of the continuum emission we detect appears to be associated
with the candidate polar ring.  The high 20\,cm flux suggests a
starburst (see Section~4).  The ``polar ring'' in this galaxy appears to
be incomplete, and the optical morphology of the incomplete ring suggests
that NGC~6286 is currently interacting with its close companion, NGC~6285
(PRC).

\begin{center}
  {\it\bf PRC C-73} (A~2358-2604)
\end{center}

These two point sources correspond to the optical positions of the two
brightest peaks at optical wavelengths (see Figure~6), which may be
two distinct galaxies in the process of merging.  This morphology is
consistent with the interpretation in the PRC, that the apparent polar
ring feature in this candidate PRG is actually tidal debris oriented so
that it resembles a polar ring in projection.  The difference between a
tidal tail that looks like a polar ring, and an actual polar ring, may
simply be a matter of the time at which one observes the interaction.
Emission from both faint point sources was included in calculations of
total radio flux and the radio spectral index for this object
(see Table~2).


\section{Star Formation vs. Nuclear Activity}
\label{SF vs AGN}

As discussed in Section~3, the radio morphology of the extended sources
in our sample is consistent with relativistic electrons from recent
supernovae as the source of most of the radio continuum emission.
To determine whether this is true for most of the galaxies in our sample,
we can examine the radio/FIR correlation and the radio spectral indices
for the candidate PRGs.

For the 17 galaxies with known fluxes at both 20\,cm and 60\,\micron, we
have plotted the radio versus the FIR fluxes in Figure~7a.  As discussed
in Section~2, it is unlikely that any of the sources we did not detect
contain strong radio sources --- our failure to detect them at 20~cm
is consistent with the FIR/radio correlation for star-forming galaxies.
None of the sources in our sample lies significantly outside the envelope
of the correlation for star-forming galaxies (from WH91), and most lie
well within it.  This is not true for early-type galaxies without polar
rings; a large fraction of these ``normal'' galaxies have significant
radio excesses over the FIR/radio correlation shown in Figure~7a, pointing
to the presence of AGN's (\eg, Reddy \&\ Yun 2004; WH91).  In our sample,
all five of the extended sources have most of their radio power in a
diffuse component, and none of the detected sources has a significant
radio excess.

It is unclear, however, whether this difference between our result
and that for non-PRG's is due to the presence of a polar ring, or to
the proportion of elliptical to lenticular galaxies.  WH91 find that
elliptical galaxies are much more likely to have a radio excess than
lenticular galaxies, and most confirmed polar rings tend to be around
lenticular galaxies rather than elliptical galaxies (PRC; van~Driel
\etal\ 2000).  However, the presence of the polar ring makes the central
galaxy difficult to classify without detailed study.

The radio spectral indices for the twelve candidate PRG's detected at
both 20\,cm and 6\,cm are shown as a histogram in Figure~7b.  The errors
for these calculated spectral indices were calculated using the 5-$\sigma$
uncertainties in the total fluxes with standard formal error calculations,
and are shown in Column~9 of Table~2.  With two exceptions (PRC~A-03 \&\
B-17), all errors are $<$\,0.1 in spectral index.

Taking S$_{\nu}$\,$\propto$\,$\nu^{\alpha}$, Figure~7b shows a slight
excess of ``steep-spectrum'' objects ($\alpha\,<\,-0.5$), compared to
``flat-spectrum'' objects ($\alpha\,>\,-0.5$).  Steep-spectrum sources
of synchrotron emission are optically thin, and generally correspond
either to regions of star formation, or the extended radio lobes of
an AGN.  We see no evidence for extended radio lobes in these PRG's.
Synchrotron emission from the ``core'' of an active nucleus is often
optically thick, so that the source has a flat spectrum.  The only one
of the flat-spectrum objects that is extended (PRC~B-21; see Figure~2)
does not have the core-jet morphology typical of an AGN.  Thus, no more
than four of our sources have the morphology and spectral indices expected
if AGN's contribute significantly to the radio emission.

Assuming, therefore, that all of the 20\,cm emission in these galaxies
is due to star formation, rather than AGN's, we have used the formula of
\citet{Co92} to calculate the rate at which massive ($M$\,$>$\,5\,\Msun)
stars are formed from the radio continuum emission (see Figure~7c).
Interestingly, the category ``A'' and ``B'' galaxies (confirmed
and probable PRG's; shown as open circles in Figure~7c) have lower
star-formation rates, on average, than the category ``C'' galaxies
(possible PRG's; shown as filled circles in Figure~7c).  None of these
implied star-formation rates are high, compared to normal disk galaxies.
For those galaxies with known \HI\ masses (see Table~2), we use these
star-formation rates to compute gas-consumption timescales, assuming
a Salpeter initial mass function truncated at $M$\,$<$\,0.4\,\Msun\
(Figure~7d).  Only one of the galaxies in our sample, PRC~C-51, could
use up all of its gas in $\sim$\,10$^8$~years, and thus may fall into the
category of ``starburst'' galaxies.  Although the majority of the radio
emission in these candidate PRGs likely arises from recent star-formation
activity, the rates implied are not high compared to the total available
gas in the galaxies.

We conclude that polar-ring galaxies have radio continuum fluxes
typical of normal (non-polar-ring) elliptical or S0 galaxies.  The
morphology of the extended sources, the spectral indices, and FIR
fluxes of the PRG's in our sample indicate that the radio emission is
predominantly due to star-forming regions rather than nuclear
activity.  However, the rate of starbirth is modest --- all but one of
our sample galaxies have enough cool gas to fuel their star formation
for at least another gigayear.


\acknowledgements

ALC and LSS acknowledge support from the National Science Foundation
through grants AST-9320403 and AST-9803114.  The work reported here
forms part of the PhD thesis of Andrea Cox, who was an NRAO Predoctoral
Fellow while much of it was carried out.
ALC would also like to acknowledge Beth Blount (Beloit College '99), 
who contributed to this paper.
The idea for this survey was a suggestion by Peter Biermann
at the Max-Planck-Institut f{\"{u}}r Radioastronomie in Bonn, Germany;
we are grateful to Barry Clark for arranging ``filler time'' at the
VLA.
This research has made use of the NASA/IPAC Extragalactic Database
(NED) which is operated by the Jet Propulsion Laboratory, California
Institute of Technology, under contract with the National Aeronautics
and Space Administration.
Optical images in Figures 1-6 were taken from the Digitized Sky Survey
(DSS), produced at the Space Telescope Science Institute
under U.S. Government grant NAG~W-2166.


\newpage
\singlespace


\begin{deluxetable}{cllrcccl}
\tablewidth{0pt}
\tabletypesize{\scriptsize}
\tablecaption{Observed sources: Data from the Polar-Ring Catalog}
\tablehead{
   Object &                & \multicolumn{2}{c}{Position}
      & Redshift  & 60\,\micron\  
                  & 100\,\micron\               & \\
   (PRC)  & Alternate Name & \multicolumn{1}{c}{$\alpha_{\rm B1950}$} 
                           & \multicolumn{1}{c}{$\delta_{\rm B1950}$}
      & (\kms)    & Flux\tablenotemark{a}~~(Jy)
                  & Flux\tablenotemark{a}~~(Jy)  & Notes 
   }
\startdata
A-01 & A~0136-0801 & 01\,36\,26\, & $-$08\,01\,24
   & 5540\tablenotemark{b} & \nodata & \nodata 
   & \\
A-03 & NGC~2685, UGC~4666 & 08\,51\,40 & $+$58\,55\,30
   & 876 & 0.3156 & 1.6942
   & ``Helix galaxy'' \\
A-04 & UGC~7576 & 12\,25\,12 & $+$28\,58\,00
   & 7035\tablenotemark{b} & \nodata & \nodata
   & Banana warp \\
A-06 & UGC~9796, II~Zw~73 & 15\,14\,00 & $+$43\,22\,00
   & 5415\tablenotemark{b} & \nodata & \nodata
   & \\
\tablevspace{8pt}
B-01 & Arp~230, IC~51 & 00\,43\,53.6\tablenotemark{c} & $-$13\,42\,55
   & 1666 & 2.2144 & 4.6906 
   & Has shells \\
B-03 & IC~1689 & 01\,20\,57 & $+$32\,47\,28
   & 4567 & \nodata & \nodata 
   & Confirmed PRG\tablenotemark{d} \\
B-09 & UGC~5119 & 09\,34\,08.1 & $+$38\,19\,01
   & 6037\tablenotemark{e} & \nodata & \nodata
   & \\
B-10 & A~0950-2234 & 09\,50\,35.2 & $-$22\,34\,27
   & 14692 & \nodata & \nodata
   & \\
B-11 & UGC~5600 & 10\,19\,17.2 & $+$78\,52\,52
   & 2769\tablenotemark{f} & 3.4639 & 4.9607
   & Inner ring?\tablenotemark{g} \\
B-12 & ESO~503-G17 & 11\,24\,24 & $-$27\,25\,48
   & 10313\tablenotemark{f} & \nodata & \nodata 
   & \\
B-16 & NGC~5122 & 13\,21\,36.8 & $-$10\,23\,39
   & 2855\tablenotemark{b} & \nodata & \nodata
   & Confirmed PRG\tablenotemark{h} \\
B-17 & UGC~9562, II~Zw~71 & 14\,49\,12 & $+$35\,45\,00
   & 1255\tablenotemark{b} & \nodata & \nodata 
   & Confirmed PRG\tablenotemark{b} \\
B-20 & A~2135-2132 & 21\,35\,31.1 & $-$21\,32\,41
   & \nodata & \nodata & \nodata
   & \\
B-21 & ESO~603-G21 & 22\,48\,40.7\tablenotemark{i} & $-$20\,30\,47
   & 3180\tablenotemark{f} & 1.4535 & 2.8893
   & Severely warped \\
B-24 & A~2333-1637 & 23\,33\,05 & $-$16\,37\,14
   & \nodata & \nodata & \nodata
   & \\
\tablevspace{8pt}
C-02 & A~0017+2212 & 00\,17\,16.7 & $+$22\,12\,00
   & \nodata & \nodata & \nodata
   & \\
C-03 & ESO~474-G26 & 00\,44\,40 & $-$24\,38\,36
   & 16246 & 0.9131 & 1.7967
   & Inner \&\ outer rings? \\
C-04 & A~0051-1323 & 00\,51\,00.2 & $-$13\,23\,04
   & 10876\tablenotemark{j} & \nodata & \nodata
   & \\
C-05 & AM~0051-234 & 00\,51\,53.5 & $-$23\,49\,26
   & 20156 & 0.4437 & 1.3828
   & \\
C-06 & NGC~304 & 00\,53\,24 & $+$23\,51\,00
   & 4991 & \nodata & \nodata
   & \\
C-09 & NGC~442 & 01\,12\,05 & $-$01\,17\,48
   & 5629\tablenotemark{f} & 0.1893 & 1.54
   & \\
C-12 & UGC~1198, VII~Zw~3 & 01\,40\,58.2 & $+$85\,00\,38
   & 1149\tablenotemark{f} & 2.58 & 3.22 
   & \\
C-24 & UGC~4261 & 08\,07\,40.2 & $+$36\,58\,38
   & 6446\tablenotemark{e} & 1.1426 & 1.56
   & \\
C-25 & UGC~4323 & 08\,15\,36.3 & $+$67\,08\,20
   & 4061\tablenotemark{e} & 0.1295 & (1.1583)
   & \\
C-27 & UGC~4385 & 08\,21\,04.0 & $+$14\,54\,49
   & 1954\tablenotemark{f} & \nodata & \nodata
   & \\
C-28 & NGC~2748 & 09\,08\,02.6\tablenotemark{k} & $+$76\,40\,53
   & 1476 & 7.0358 & 17.982
   & \\
C-29 & NGC~2865 & 09\,50\,35.2 & $-$22\,34\,27
   & 14692 & \nodata & \nodata
   & \\
\tablebreak
C-30 & UGC~5101 & 09\,32\,05.1 & $+$61\,34\,33
   & 11945\tablenotemark{l} & 11.542 & 20.226
   & \\
C-32 & IC~575 & 09\,52\,04 & $-$06\,37\,00
   & 5973\tablenotemark{m} & 0.2716 & 0.9833
   & \\
C-35 & NGC~3414 & 10\,48\,31.8 & $+$28\,14\,24
   & 1360 & 0.1907 & 0.6132
   & \\
C-37 & UGC~6182 & 11\,05\,08.3 & $+$53\,53\,13
   & 1255 & \nodata & \nodata
   & \\
C-49 & NGC~6028 & 15\,59\,15.7 & $+$19\,29\,52
   & 4475 & \nodata & \nodata
   & \\
C-50 & UGC~10205 & 16\,04\,36 & $+$30\,14\,00
   & 6605\tablenotemark{l} & 0.3939 & 1.5383
   & \\
C-51 & NGC~6286, Arp 293 
   & 16\,57\,44.7\tablenotemark{n} & $+$59\,00\,40
   & 5501\tablenotemark{e} & 7.8778 & 22.595
   & Interacting; see Fig. 5 \\
C-54 & A~2027-2335 & 20\,27\,30.8 & $-$23\,35\,32
   & \nodata & 0.2364 & (0.5228)
   & \\
C-60 & ESO~464-G31 & 21\,15\,25.4 & $-$27\,33\,35
   & 6529\tablenotemark{f} & \nodata & \nodata
   & \\
C-66 & A~2150-1707 & 21\,50\,20.9 & $-$17\,07\,50
   & \nodata & \nodata & \nodata
   & \\
C-69 & NGC~7468 & 23\,00\,30.5 & $+$16\,20\,06
   & 2072\tablenotemark{f} & 1.2503 & 1.4890
   & \\
C-71 & ZGC~2315+03 & 23\,15\,42 & $+$03\,54\,00
   & 18770 & \nodata & \nodata
   & Projection effect?\tablenotemark{o}\\
C-73 & A~2358-2604 & 23\,58\,41.9 & $-$26\,04\,48
   & 15364\tablenotemark{p}  & 0.1497 & (0.3849)
   & ``Ring'' $=$ tidal debris? \\
\enddata
\tablenotetext{a}{
   Most 60\,\micron\ \&\ 100\,\micron\ fluxes come from the IRAS
   Faint-Source Catalog (FSC).  Fluxes listed to only two decimal
   places come from the Point-Source Catalog (PSC).
   }
\tablenotetext{b}{Cox \etal\ 2001}
\tablenotetext{c}{Coziol \etal\ 1993}
\tablenotetext{d}{Sil'chenko 1998}
\tablenotetext{e}{Reshetnikov \&\ Combes 1994}
\tablenotetext{f}{Richter, Sackett, \&\ Sparke 1994}
\tablenotetext{g}{Karataeva \etal\ 2001}
\tablenotetext{h}{Cox 1996}
\tablenotetext{i}{Loveday 1996}
\tablenotetext{j}{Huchra \etal\ 1993}
\tablenotetext{k}{Dressel \&\ Condon 1976}
\tablenotetext{l}{de~Vauclouleurs \etal\ 1991}
\tablenotetext{m}{Bottinelli \etal\ 1993}
\tablenotetext{n}{Russell \etal\ 1990}
\tablenotetext{o}{Karataeva, Hagen-Thorn, \&\ Yakovleva 2000}
\tablenotetext{p}{Redshift from Ratcliffe \etal\ 1998}
\end{deluxetable}

\begin{deluxetable}{ccccccl}
\setlength{\tabcolsep}{0.5\tabcolsep}
\tablewidth{400pt}
\tablecaption{Radio fluxes and spectral indices}
\tablehead{
   Object & $\int{S_{\rm 21\,cm}\,dv}$\tablenotemark{a} 
          & M(\HI)
          & S$_{\rm 20\,cm}$
          & S$_{\rm 6\,cm}$ 
          & $\alpha_{\rm 20\,cm, 6\,cm}$   
          & \\
   (PRC)  & (Jy km s$^{-1}$) 
          & (10$^9$ h$^{-2}$ \Msun)
          & (mJy)  
          & (mJy)
          & (S$_{\nu}\,\propto\,\nu^{\alpha}$) 
          & Note\tablenotemark{b}
   }
\startdata
%
%
A-01 & 2.2\tablenotemark{c} 
           & 1.6 & $<$~0.55 & \nodata & \nodata & \\
A-03 & 34\tablenotemark{d}
           & 0.6 & 0.70 $\pm$ 0.06 & 0.51 $\pm$ 0.08  
           & $-$0.26 $\pm$ 0.21 & U \\
A-04 & 2.3\tablenotemark{c}
           & 2.7 & $<$~0.55 & \nodata & \nodata & \\ 
A-06 & 3.7\tablenotemark{c}
           & 2.6 & $<$~0.55 & \nodata & \nodata & \\
%
%
B-01 & $\sim$\,9 & $\sim$\,0.6
           & 24.0 $\pm$ 0.9 & 10.7 $\pm$ 0.4 & $-$0.67 $\pm$ 0.06 & E \\
B-03 &     &     &  $<$~0.55 & \nodata & \nodata & \\
B-09 & 2.4 & 2.1 &  $<$~0.55 & \nodata & \nodata & \\
B-10 &     &     &  $<$~0.55 & \nodata & \nodata & \\
B-11 & 11  & 2.0 & 10.3 $\pm$ 0.3 &  5.5 $\pm$ 0.2 & $-$0.52 $\pm$ 0.06 & U \\
B-12 & 1.5 & 3.8 &  $<$~0.55 & \nodata & \nodata & \\
B-16 & 3.4\tablenotemark{c}
           & 0.7 &  $<$~0.45\tablenotemark{c}  &  $<$~0.55 & \nodata & \\
B-17 & 6.2\tablenotemark{e}
           & 0.2 &  1.2 $\pm$ 0.20 & 0.3 $\pm$ 0.09 
           & $-$1.15 $\pm$ 0.40 & U\tablenotemark{f} \\
     &     &     &  4.5 $\pm$ 0.2 & 1.3 $\pm$ 0.1 & $-$1.03  $\pm$ 0.11 & U \\
     &     &     &  0.7 $\pm$ 0.2 & 0.9 $\pm$ 0.09 & $+$0.21 $\pm$ 0.30 & U \\
B-20 &     &     &  $<$~0.55 & \nodata & \nodata & \\
B-21 & 14\tablenotemark{g}  
           & 3.3 &  9.1 $\pm$ 0.5 & 5.1 $\pm$ 0.2 & $-$0.48 $\pm$ 0.08 & E \\
B-24 &     &     &  $<$~0.50 & \nodata & \nodata & \\
%
%
C-02 &     &     &  $<$~0.50 & \nodata & \nodata & \\
C-03 &     &     &  6.5 $\pm$ 0.2 & 2.5 $\pm$ 0.1 & $-$0.79 $\pm$ 0.06 & U \\
C-04 &     &     &  0.8 $\pm$ 0.08 & 0.9 $\pm$ 0.05 & $+$0.10 $\pm$ 0.13 & U \\
C-05 &     &     & 13.4 $\pm$ 0.4 & 4.6 $\pm$ 0.2 & $-$0.89 $\pm$ 0.05 & U \\
C-06 &     &     &  1.0 $\pm$ 0.1 & \nodata & \nodata & B \\
C-09 & 1.5 & 1.1 &  $<$~0.55 &  $<$~0.40 & \nodata & \\
C-12 & 1.0 & 0.03 &  4.9 $\pm$ 0.2 & 4.3 $\pm$ 0.1 & $-$0.11 $\pm$ 0.06 & B \\
C-24 & 4.6 & 4.5 & 11.10 $\pm$ 0.4 & \nodata & \nodata & E\tablenotemark{h} \\
     &    &    & 0.74 $\pm$ 0.07 & \nodata & \nodata & U \\
C-25 & 3.6 & 1.4 &  1.0 $\pm$ 0.1 & \nodata & \nodata & U \\
C-27 & 7.0 & 0.6 & $<$~0.55 & \nodata & \nodata & \\
C-28 & 30\tablenotemark{i} & 1.5 & 38.9 $\pm$ 2.0 & \nodata & \nodata & E \\
C-29 & 2.0 & 10.2 & $<$~0.55 & \nodata & \nodata & \\
\tablebreak
%
%
C-30\tablenotemark{j} 
     &     &     & 148.7 $\pm$ 0.28 & 65.4 $\pm$ 0.58 & $-$0.68 $\pm$ 0.01 & B \\
C-32 & 1.6 & 1.3 & 4.6 $\pm$ 0.2 & \nodata & \nodata & U \\
C-35 & 1.0 & 0.04 & 3.5 $\pm$ 0.1 & \nodata & \nodata & U \\
C-37 & 10  & 0.4 & $<$~0.50 & \nodata & \nodata & \\
C-49 & 2.0 & 0.9 & $<$~0.50 & \nodata & \nodata & \\
C-50 & 2.0 & 2.1 & 3.6 $\pm$ 0.1 & \nodata & \nodata & U \\
C-51 & 1.0 & 0.7 & 143 $\pm$ 4 & \nodata & \nodata & E\tablenotemark{k} \\
     &     &     & 9.6 $\pm$ 0.4 & \nodata & \nodata & B \\
C-54 &     &     & $<$~0.55 & $<$~0.45 & \nodata & \\
C-60 & 5.0 & 5.0 &  0.9 $\pm$ 0.08 & 1.1 $\pm$ 0.06 & $+$0.17 $\pm$ 0.12 & U \\
C-66 &     &     & $<$~0.55 & \nodata & \nodata & \\
C-69 & 10  & 1.0 & 8.9 $\pm$ 0.3 &  4.1 $\pm$ 0.1 & $-$0.64 $\pm$ 0.06 & B \\
C-71 &     &     & $<$~0.50 & \nodata & \nodata & \\
C-73 &     &     &  2.1 $\pm$ 0.2 & 0.56 $\pm$  0.06 & $-$1.10 $\pm$ 0.05
                                                     & U\tablenotemark{l} \\
     &     &     &  2.6 $\pm$ 0.1 & 0.37 $\pm$ 0.04 & $-$1.62 $\pm$ 0.14 & U \\
     &     &     &  7.8 $\pm$ 0.3 & 2.49 $\pm$ 0.1 & $-$0.95 $\pm$ 0.07 & B \\
\enddata
\tablenotetext{a}{
   Unless otherwise noted, 21\,cm line integrals are taken from 
   van~Driel \etal\ 2000.
   }
\tablenotetext{b}{
   ``U'' means that the source was unresolved in this survey, ``B'' that
   it was barely resolved, and ``E'' that it was extended.  All sources
   were observed with $\simeq\,5\arcsec$ resolution.
   }
\tablenotetext{c}{
   Cox 1996
   }
\tablenotetext{d}{
   Richter, Sackett, \&\ Sparke 1994
   }
\tablenotetext{e}{
   Cox \etal\ 2001
   }
\tablenotetext{f}{
   The first source is at the position of PRC~B-17 (II~Zw~71), and the
   second is at the position of II~Zw~70, the interacting companion to
   PRC~B-17.  The third source, possibly unrelated, is placed near the
   eastern extension of the \HI\ gas we observe in PRC~B-17 (see Cox
   \etal\ 2001).
   }
\tablenotetext{g}{
   PRC~B-21 has a close companion with a similar redshift; some of the \HI\
   is probably associated with the companion.
   }
\tablenotetext{h}{
   The first source is at the position of PRC~C-24 (UGC~4261), and 
   the second source is assumed to be unrelated (see Figure~3).
   }
\tablenotetext{i}{
   Kamphuis, Sijbring, \&\ van~Albada 1996
   }
\tablenotetext{j}{
   The data for this object, including uncertainty estimates, are 
   from Crawford \etal\ (1996).
   }
\tablenotetext{k}{
   The first source is at the position of PRC~C-51 (NGC~6286), and
   the second source is at the position of its companion, NGC~6285.
   }
\tablenotetext{l}{
   Two faint point sources are observed at the position of PRC C-73
   (see Figure~6);  the sum of the flux from both of these sources is 
   listed on the first line.  The two brighter, more southern
   sources are probably unrelated.
   }
\end{deluxetable}


\clearpage
\begin{figure}
\includegraphics[width=3.0in]{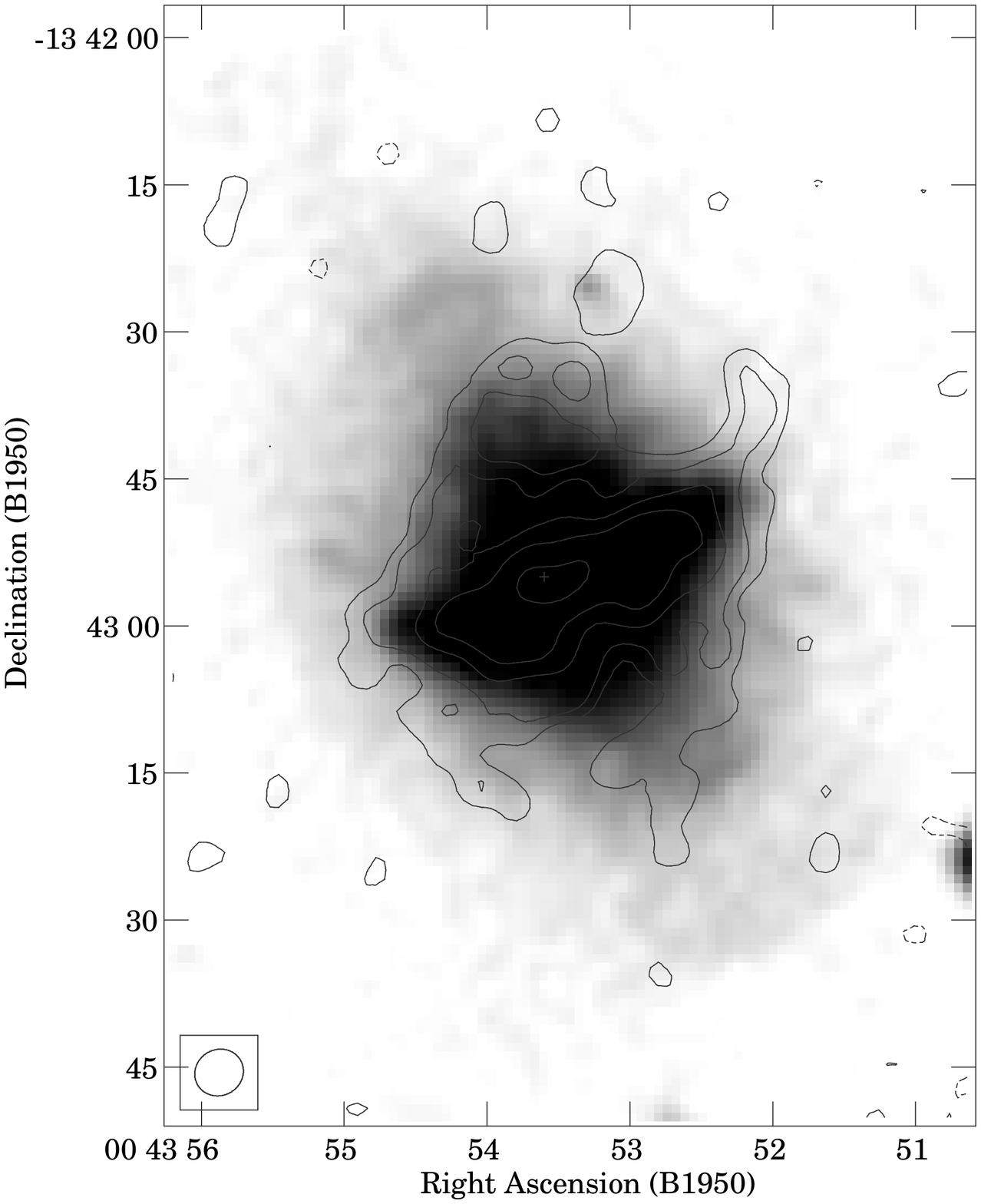}
\includegraphics[width=3.0in]{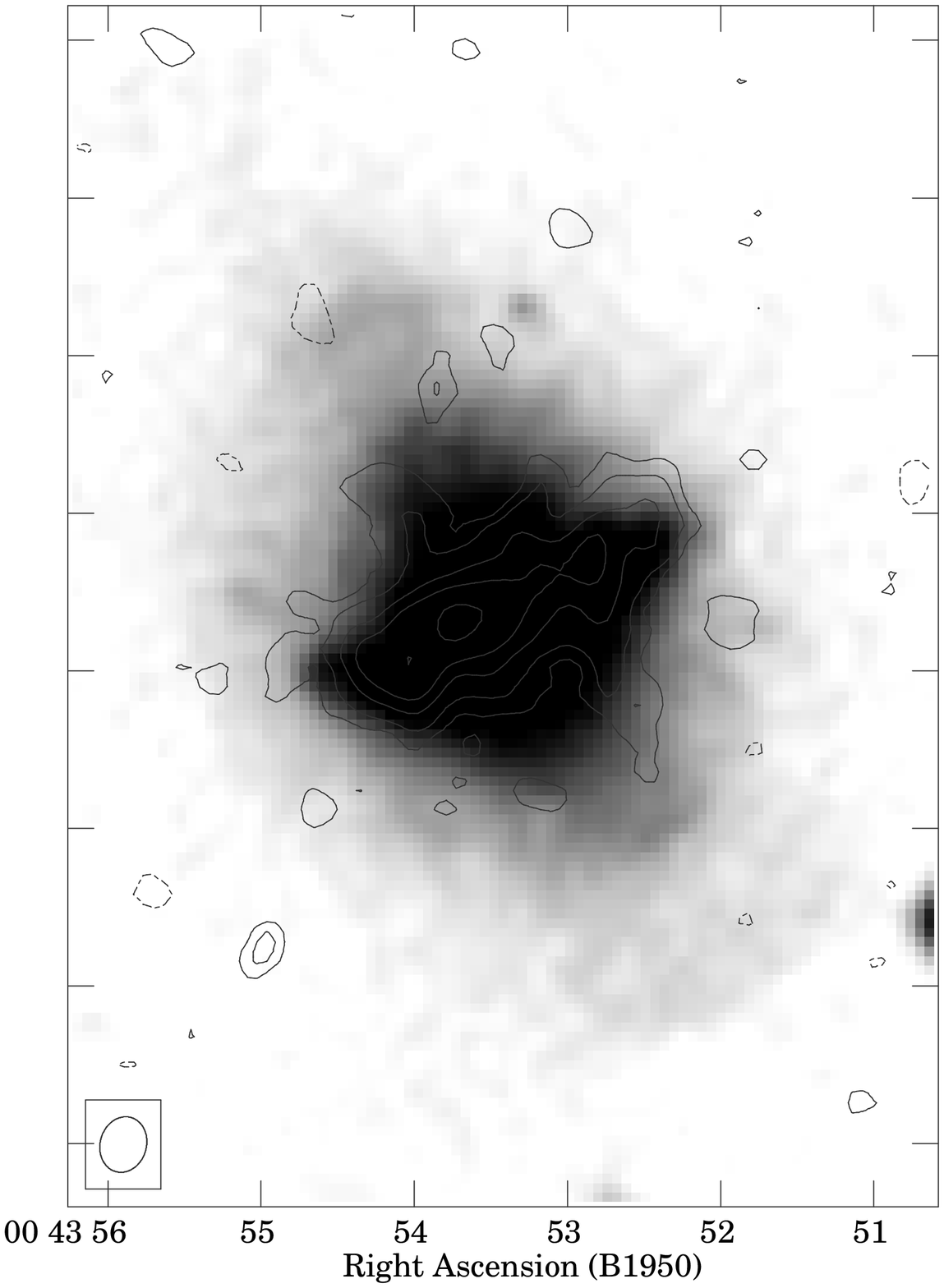}
\caption[PRC B-01:  Contour plots of 20\,cm \&\ 6\,cm emission]{
   Contour plots of the 20\,cm ({\bf left}) and 6\,cm ({\bf right})
   emission in PRC~B-01 (IC~51, Arp~230), overlaid on an optical image
   from the STScI Digitized Sky Survey (DSS).  The synthesized beam is
   shown in the lower right-hand corner.  Our 5\arcsec\ resolution
   corresponds to 0.4~$h^{-1}$~kpc at the distance of this galaxy.
   Negative contours are dashed; contours are at $-$2.5, 2.5, 4, 7, 10,
   15, 20, \&\ 25-$\sigma$ ($\sigma$\,$=$0.07~mJy~beam$^{-1}$ at
   20\,cm; $\sigma$\,$=$0.05~mJy~beam$^{-1}$ at 6\,cm).
   }
\end{figure}

\clearpage
\begin{figure}
\begin{center}
\includegraphics[angle=-90, width=4in]{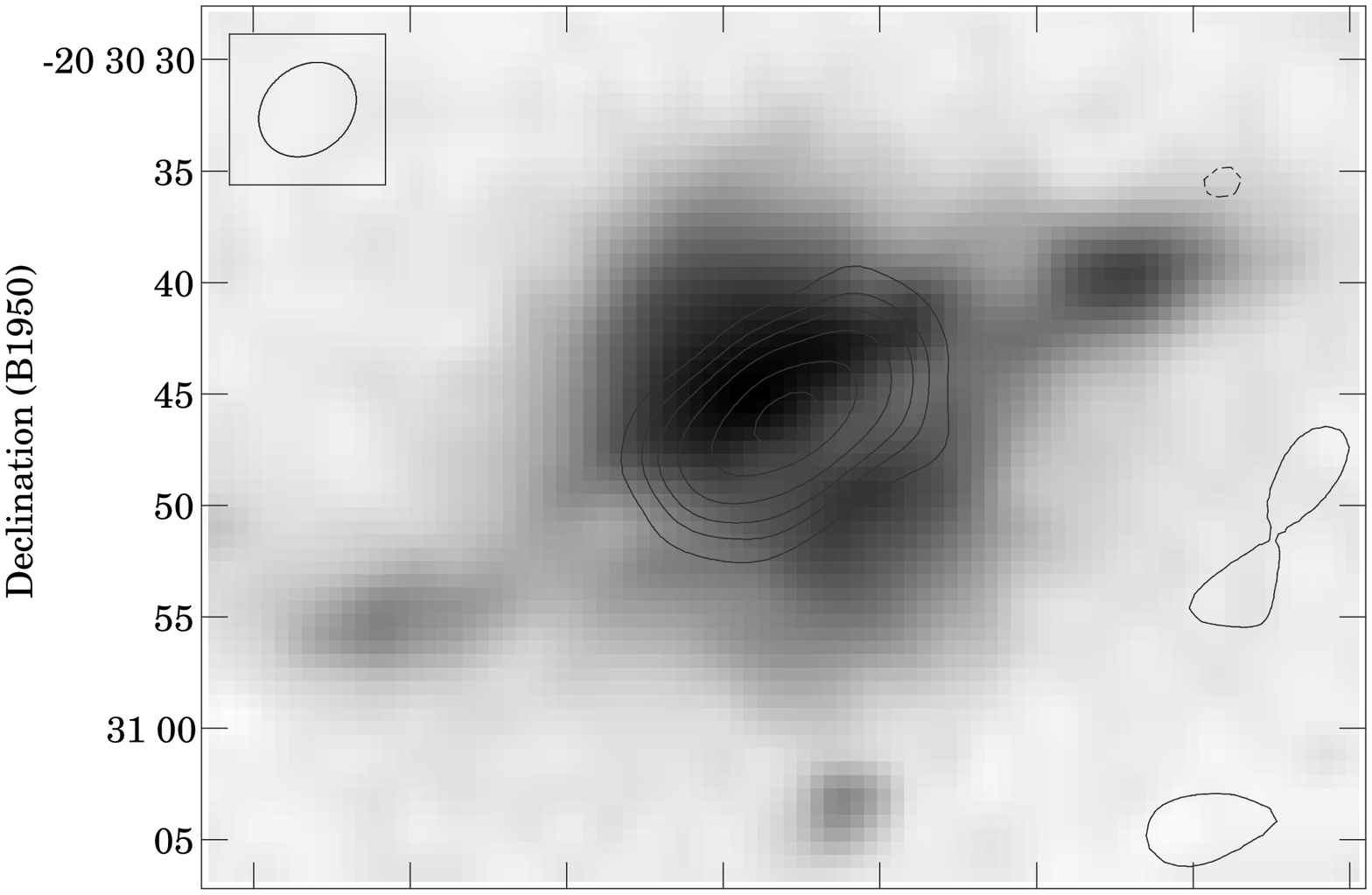}\\
\includegraphics[angle=-90, width=4in]{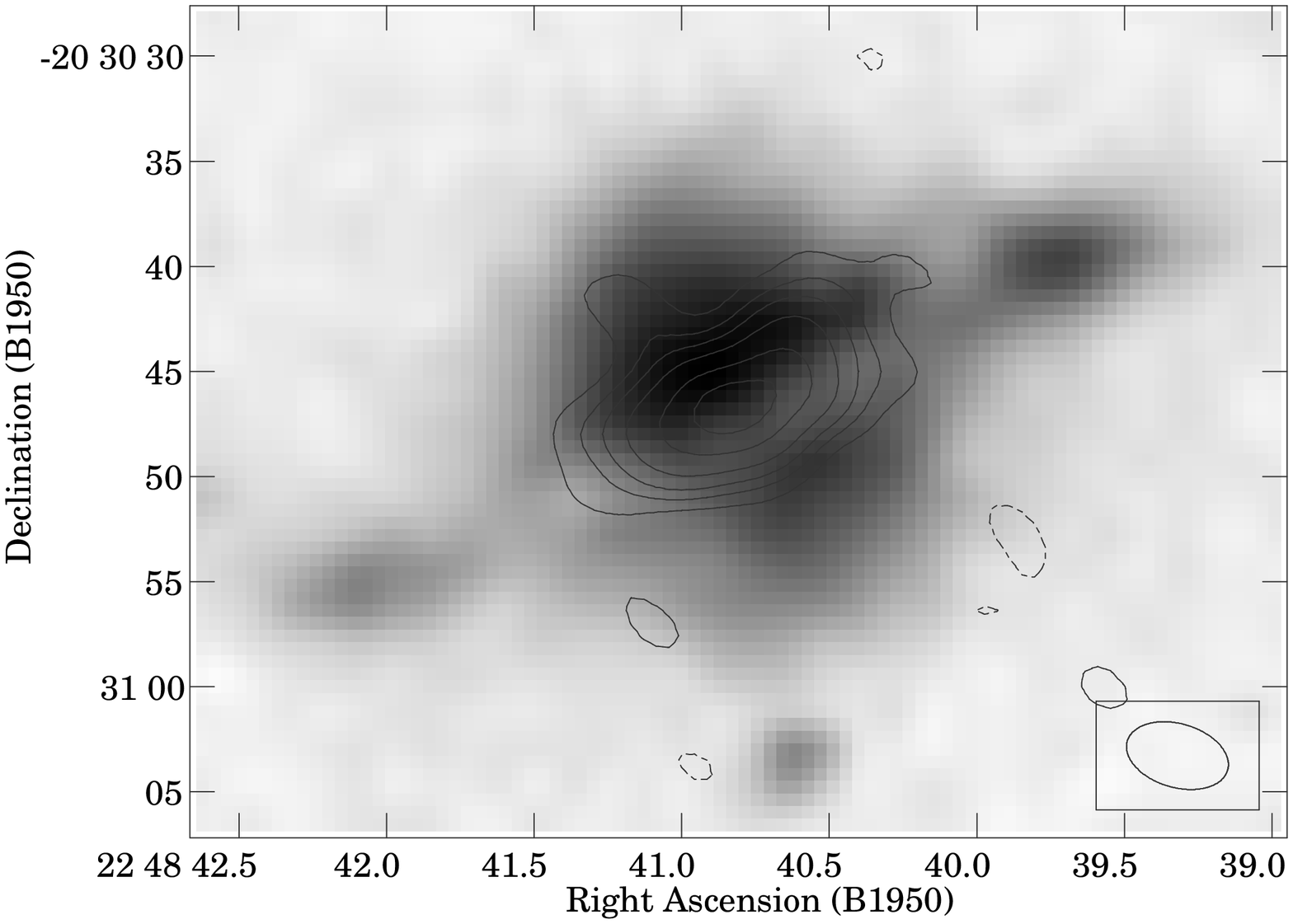}
\end{center}
\caption[PRC~B-21: Contour plots of 20\,cm \&\ 6\,cm emission]{
   Contour plots of the 20\,cm ({\bf top}) and 6\,cm ({\bf bottom})
   emission in PRC~B-21 (ESO~603-G21), overlaid on an optical image
   from the DSS.  The synthesized beams are shown in the upper left and
   lower right corners, respectively.  Our 5\arcsec\ resolution
   corresponds to 0.8~$h^{-1}$~kpc at the distance of this galaxy.
   Negative contours are dashed; contours are at $-$3, 3, 5, 7, 10, 15,
   20, \&\ 25-$\sigma$ ($\sigma$\,$=$0.13~mJy~beam$^{-1}$ at 20\,cm;
   $\sigma$\,$=$0.05~mJy~beam$^{-1}$ at 6\,cm).
   }
\end{figure}

\clearpage
\begin{figure}
\includegraphics[angle=-90, width=\textwidth]{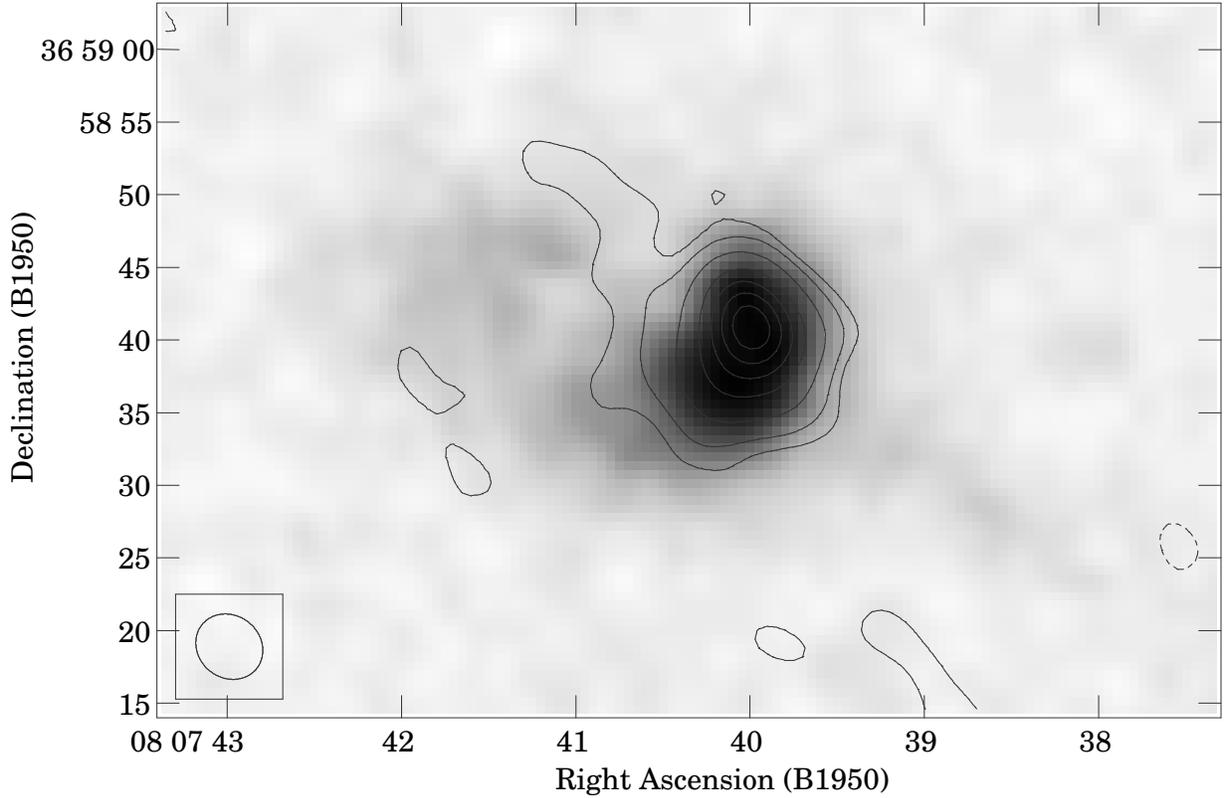}
\caption[PRC~C-24: Contour plot of 20\,cm emission]{
   Contour plot of the 20\,cm emission in PRC~C-24 (UGC~4261), overlaid
   on an optical image from the DSS.  The synthesized beam is shown in
   the lower left corner.  Our 5\arcsec\ resolution corresponds to
   1.6~$h^{-1}$~kpc at the distance of this galaxy.  Negative contours
   are dashed; contours are at $-$3, 3, 5, 10, 20, 30, 40,
   \&\ 50-$\sigma$ ($\sigma$\,$=$\,0.07~mJy~beam$^{-1}$).  The peak
   flux density at the bright central core is 4.1~mJy~beam$^{-1}$.
   }
\end{figure}

\begin{figure}		
\plotone{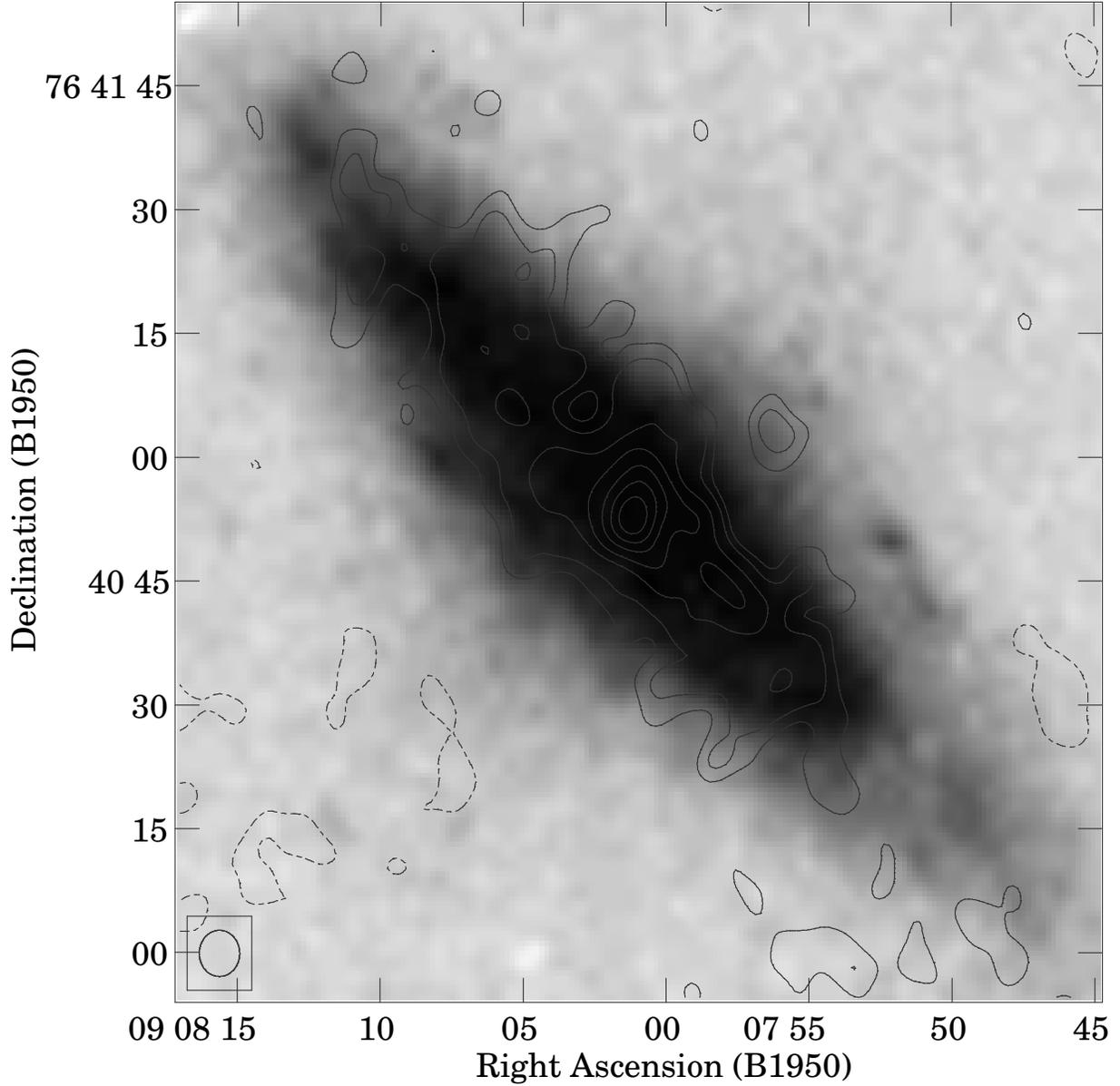}
\caption[PRC~C-28: Contour plot of 20\,cm emission]{
   Contour plot of the 20\,cm emission in PRC~C-28 (NGC~2748), overlaid
   on an optical image from the DSS.  The synthesized beam is shown in
   the lower left corner.  Our 5\arcsec\ resolution corresponds to
   0.4~$h^{-1}$~kpc at the distance of this galaxy.  Negative contours
   are dashed; contours are at $-$2.5, 2.5, 4, 7, 10, 14, 18,
   \&\ 22-$\sigma$ ($\sigma$\,$=$\,0.10~mJy~beam$^{-1}$).
   }
\end{figure}

\clearpage
\begin{figure}
\plotone{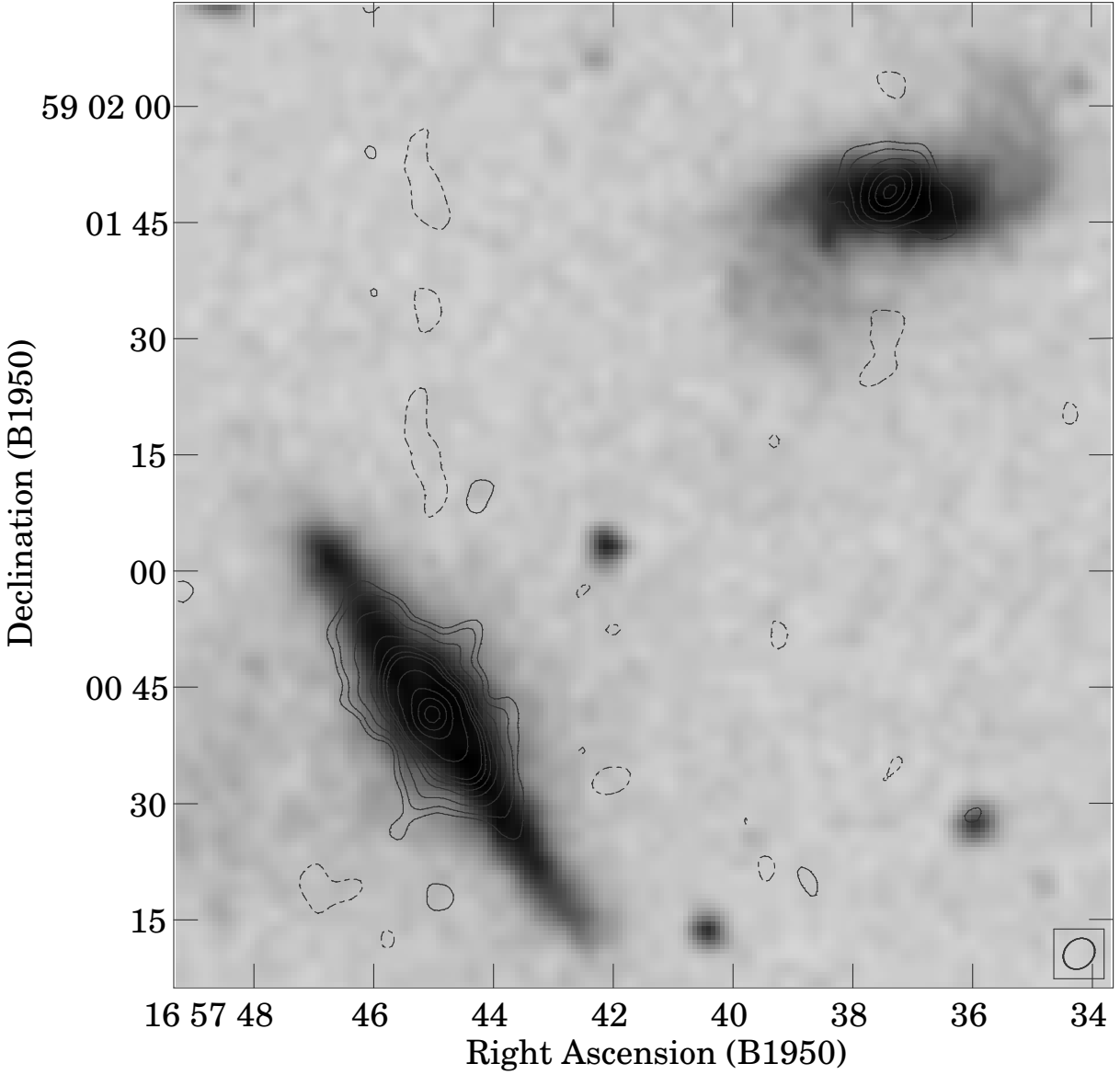}
\caption[PRC~C-51: Contour plot of 20\,cm emission]{
   Contour plot of the 20\,cm emission in PRC~C-51 (NGC~6286), overlaid
   on an optical image from the DSS.  The galaxy to the northwest of
   C-51 is its companion galaxy, NGC~6285.  The synthesized beam is
   shown in the lower right corner.  Our 5\arcsec\ resolution
   corresponds to 1.3~$h^{-1}$~kpc at the distance of this galaxy.
   Negative contours are dashed; contours are at $-$3, 3, 5, 10, 20,
   30, 40, 50, 100, 200, 300, \&\ 400-$\sigma$
   ($\sigma$\,$=$\,0.11~mJy~beam$^{-1}$).  The approximately linear
   negative contours are artefacts of the imaging \&\ deconvolution
   process.
   }
\end{figure}

\clearpage
\begin{figure}
\plottwo{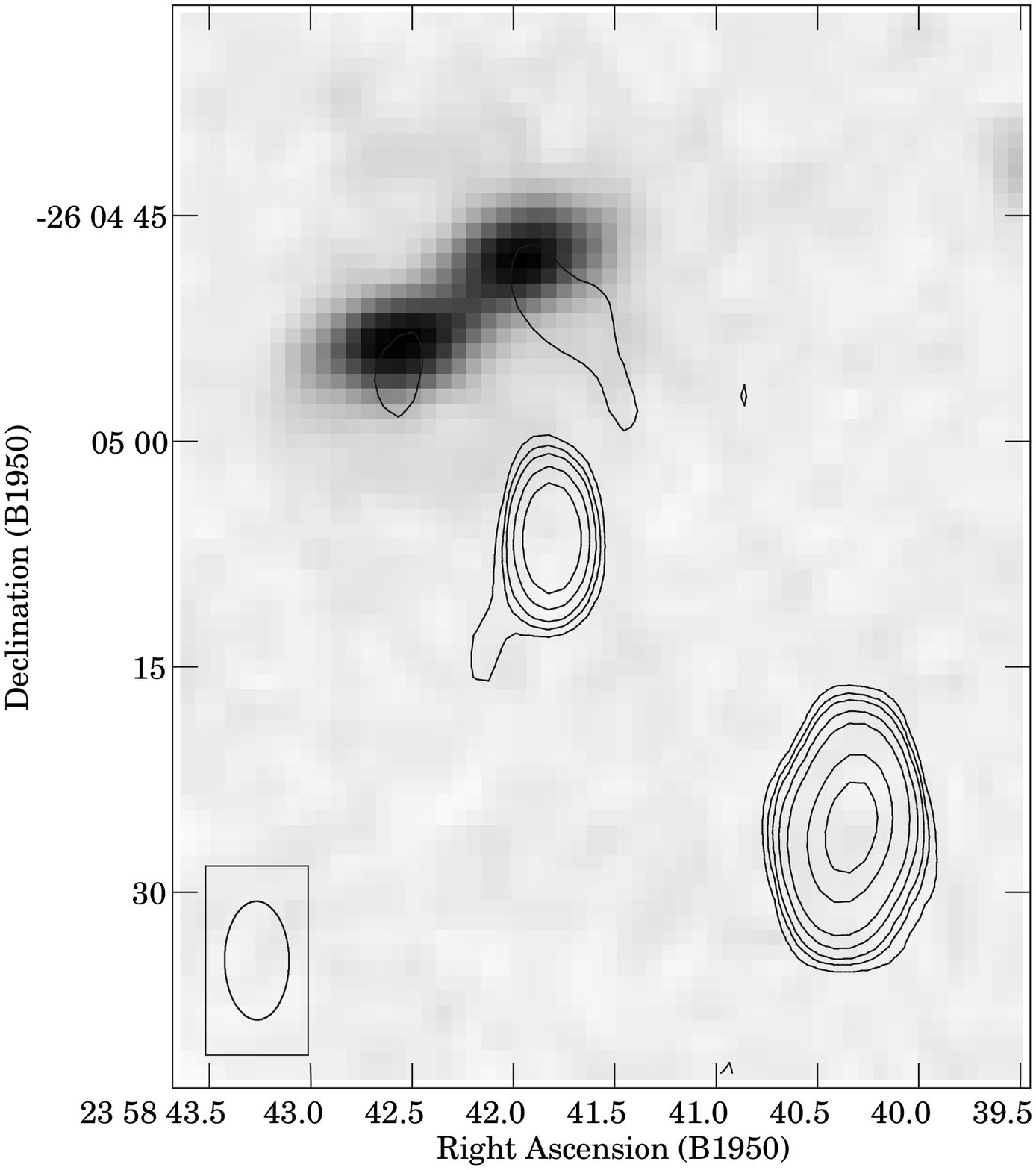}{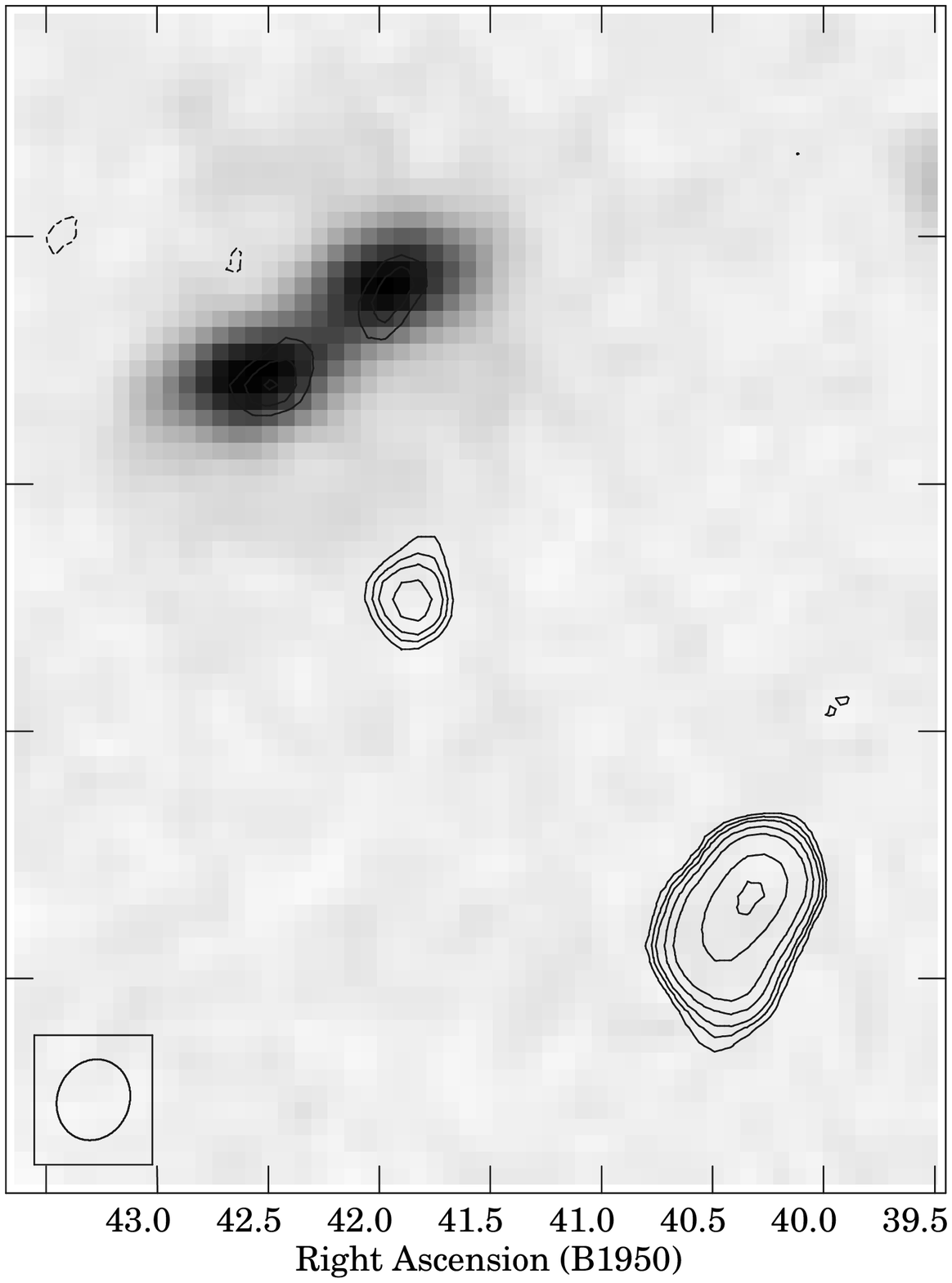}
\caption[PRC~C-73:  Contour plot of 20\,cm \&\ 6\,cm emission]{
   Contour plots of the 20\,cm ({\bf left}) and 6\,cm ({\bf right})
   emission in PRC~C-73 (A~2358-2604), overlaid on an optical image
   from the DSS.  From both the optical morphology and the presence of
   two separate radio continuum sources, it is probable that this PRG
   candidate is a superposition of two (possibly interacting)
   galaxies.  The two brighter continuum sources to the south are
   probably unrelated.  The synthesized beams are shown in the lower
   left corners.  Our 5\arcsec\ resolution corresponds to
   3.7~$h^{-1}$~kpc at the distance of this galaxy.  Negative contours
   are dashed; contours are at $-$3, 3, 4, 5, 7, 10, 20,
   \&\ 30-$\sigma$ ($\sigma$\,$=$\,0.11~mJy~beam$^{-1}$ at 20\,cm;
   $\sigma$\,$=$\,0.04~mJy~beam$^{-1}$ at 6\,cm).
   }
\end{figure}


\clearpage
\begin{figure}
\plotone{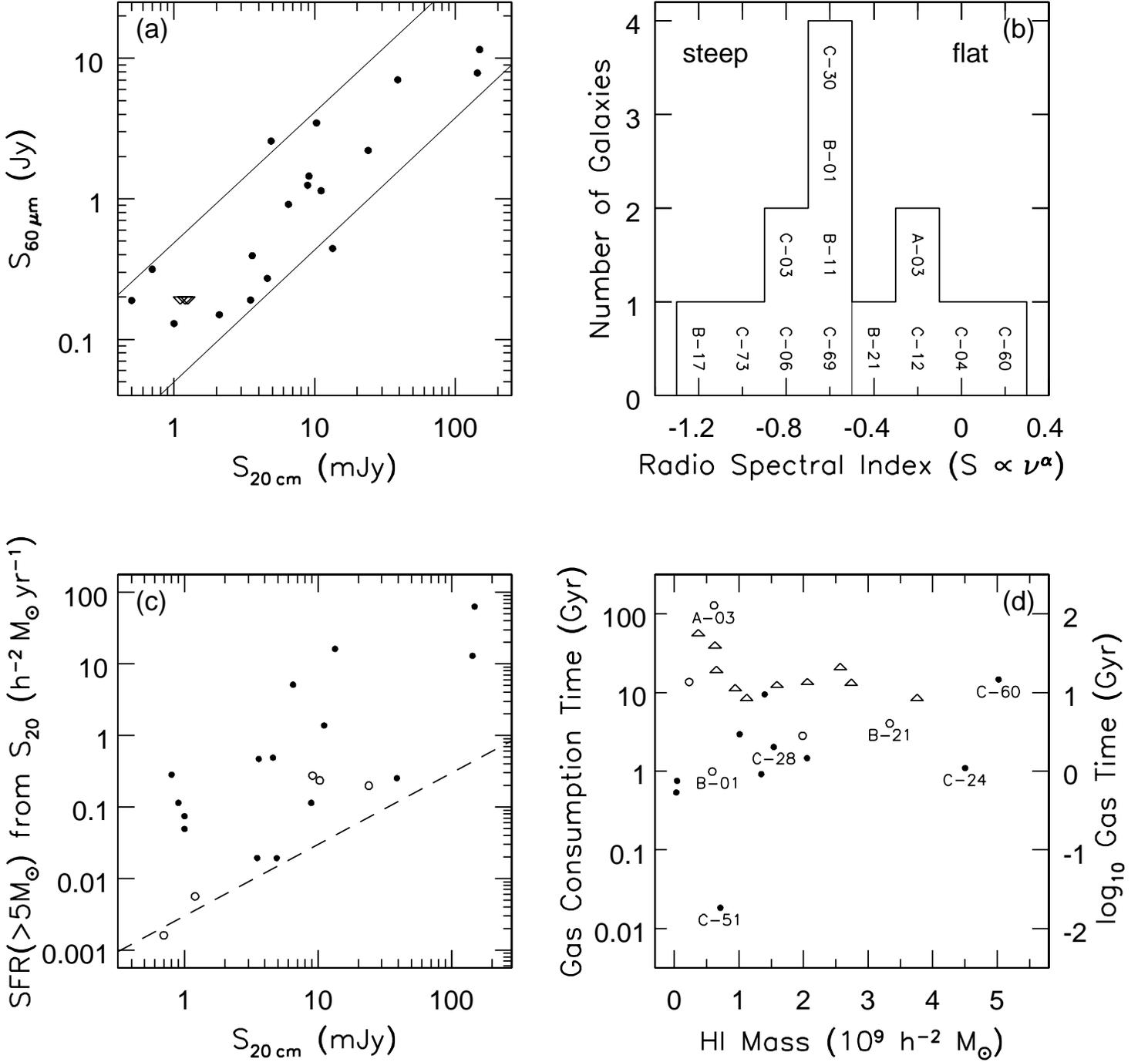}
\caption[Useful figures from the data]{
   {\bf (a)} 60\,$\mu$m FIR flux versus 20\,cm flux for
   the galaxies detected in either band.  Triangles show upper limits;
   Sloping lines show the envelope of the FIR/radio correlation for
   star-forming galaxies from Wrobel \&\ Heeschen (1991), assuming an
   average radio spectral index $\alpha$\,$=$\,$-$0.5.
   {\bf (b)} Radio spectral indices of detected PRG's at 20\,cm
   \&\ 6\,cm.  Two of these sources, B-01 \&\ B-21, are extended at
   5\arcsec\ resolution (see Figures~1 \&\ 2).  One of these, C-73, may
   be two interacting galaxies which look like a PRG in
   projection (PRC).
   {\bf (c)} Birthrate of stars more massive than 5\,\Msun, calculated
   from the 20\,cm continuum total flux.  Category A/B PRG's (open
   circles) are generally forming stars at a lower rate than category C
PRG's
   (filled circles).  The dashed line represents an object at a
   distance of 10\,$h^{-1}$\,Mpc.
   {\bf (d)} \HI\ gas-consumption timescale for PRG's with known 21\,cm
   line integrals (see Table~2).  The total SFR is estimated as three
   times the SFR\,$>$\,5\,\Msun\ (panel~c).  Open circles are category
   A/B objects and filled circles are category C objects.  Triangles
   are lower limits for galaxies with known \HI\ masses that were
   undetected at 20\,cm, assuming that the undetected emission would be
   from a single point source.  All of the
   sources with extended emission are labeled (except for PRC~C-73,
   which has no known \HI\ mass), as are the two most extreme
   objects on the plot (PRC~A-03 \&\ PRC~C-60).
   }
\end{figure}

\end{document}